\documentclass[dvips]{acta}
\usepackage{supertabular,lscape,epsfig}
\usepackage{amssymb}
\usepackage{amsmath}
\DeclareSymbolFont{ppa}{OT1}{ppl}{m}{it}
\DeclareMathSymbol{\vv}{\mathalpha}{ppa}{'166}

\newfont{\hb}{rphvb at 10pt}%bezszeryfowe pó³grube
\newfont{\hbo}{rphvbo at 10pt}%bezszeryfowe pó³grube kursywa
\newfont{\bitt}{rptmbi at 12pt}%pó³gruba kursywa (tytu³ artyku³u)
\newfont{\bits}{rptmbi at 11pt}%pó³gruba kursywa (tytu³y rozdzia³ów)

\SetPages{37}{40}

\SetVol{63}{2013}

\begin{document}

%Zwarte naglowki, jeden wiersz, appendix
\newcommand{\TabApp}[2]{\begin{center}\parbox[t]{#1}{\centerline{
  {\bf Appendix}}
  \vskip2mm
  \centerline{\small {\spaceskip 2pt plus 1pt minus 1pt T a b l e}
  \refstepcounter{table}\thetable}
  \vskip2mm
  \centerline{\footnotesize #2}}
  \vskip3mm
\end{center}}

%Zwarte naglowki, jeden wiersz
\newcommand{\TabCapp}[2]{\begin{center}\parbox[t]{#1}{\centerline{
  \small {\spaceskip 2pt plus 1pt minus 1pt T a b l e}
  \refstepcounter{table}\thetable}
  \vskip2mm
  \centerline{\footnotesize #2}}
  \vskip3mm
\end{center}}

%Zwarte naglowki, dwa wiersze
\newcommand{\TTabCap}[3]{\begin{center}\parbox[t]{#1}{\centerline{
  \small {\spaceskip 2pt plus 1pt minus 1pt T a b l e}
  \refstepcounter{table}\thetable}
  \vskip2mm
  \centerline{\footnotesize #2}
  \centerline{\footnotesize #3}}
  \vskip1mm
\end{center}}

%Zwarte naglowki, jeden wiersz, appendix
\newcommand{\MakeTableApp}[4]{\begin{table}[p]\TabApp{#2}{#3}
  \begin{center} \TableFont \begin{tabular}{#1} #4 
  \end{tabular}\end{center}\end{table}}

%Zwarte naglowki, jeden wiersz
\newcommand{\MakeTableSepp}[4]{\begin{table}[p]\TabCapp{#2}{#3}
  \begin{center} \TableFont \begin{tabular}{#1} #4 
  \end{tabular}\end{center}\end{table}}

%Zwarte naglowki, jeden wiersz
\newcommand{\MakeTableee}[4]{\begin{table}[htb]\TabCapp{#2}{#3}
  \begin{center} \TableFont \begin{tabular}{#1} #4
  \end{tabular}\end{center}\end{table}}

%Zwarte naglowki, dwa wiersze
\newcommand{\MakeTablee}[5]{\begin{table}[htb]\TTabCap{#2}{#3}{#4}
  \begin{center} \TableFont \begin{tabular}{#1} #5 
  \end{tabular}\end{center}\end{table}}

%{\it Acta Astronomica Archive}
%\parskip=0pt \itemsep=1mm \setlength{\itemsep}{0.4mm}\setlength{\parindent}{-1em} \setlength{\itemindent}{-1em} - po \begin{itemize} - wszystko
%FWHM, PSF, S/N - proste, 
%MgII, H$\alpha$
%rms, rhs, sd - kursywa
%{\sc DAOPhot}
%{\sc Fnpeaks}
%{\sf files}
%Galactic wszystko (bulge, center, plane, disk, coordinates, latitudes...)
%Cepheids
%type~ Cepheids, Population~II Cepheids
%a.u.
\newfont{\bb}{ptmbi8t at 12pt}
\newfont{\bbb}{cmbxti10}
\newfont{\bbbb}{cmbxti10 at 9pt}
\newcommand{\uprule}{\rule{0pt}{2.5ex}}
\newcommand{\douprule}{\rule[-2ex]{0pt}{4.5ex}}
\newcommand{\dorule}{\rule[-2ex]{0pt}{2ex}}
\def\thefootnote{\fnsymbol{footnote}}
\hyphenation{OGLE}

\begin{Titlepage}
\Title{The Optical Gravitational Lensing Experiment.\\
The OGLE-III Catalog of Variable Stars.\\
Type~II Cepheids in the Galactic Bulge -- Supplement\footnote{Based on
observations obtained with the 1.3-m Warsaw telescope at the Las Campanas
Observatory of the Carnegie Institution for Science.}}
\vspace*{5pt}
\Author{I.~~S~o~s~z~y~ñ~s~k~i$^1$,~~
A.~~U~d~a~l~s~k~i$^1$,~~
P.~~P~i~e~t~r~u~k~o~w~i~c~z$^1$,~~
M.\,K.~~S~z~y~m~a~ñ~s~k~i$^1$,\\
M.~~K~u~b~i~a~k$^1$,~~
G.~~P~i~e~t~r~z~y~ñ~s~k~i$^{1,2}$,~~
£.~~W~y~r~z~y~k~o~w~s~k~i$^{1,3}$,~~
K.~~U~l~a~c~z~y~k$^1$,~~\\
R.~~P~o~l~e~s~k~i$^{4,1}$~~
and~~S.~~K~o~z~³~o~w~s~k~i$^1$}
{$^1$Warsaw University Observatory, Al.~Ujazdowskie~4, 00-478~Warszawa, Poland\\
e-mail:
(soszynsk,udalski,pietruk,msz,mk,pietrzyn,kulaczyk,rpoleski,simkoz)@astrouw.edu.pl\\
$^2$Universidad de Concepción, Departamento de Astronomia, Casilla 160--C, Concepción, Chile\\
$^3$Institute of Astronomy, University of Cambridge, Madingley Road, Cambridge CB3~0HA,~UK\\ 
e-mail: wyrzykow@ast.cam.ac.uk\\
$^4$Department of Astronomy, Ohio State University, 140 W.~18th Ave., Columbus, OH~43210, USA}
\Received{March 26, 2013}
\end{Titlepage}

\Abstract{We report the discovery of additional 22 RV~Tau stars located
in the OGLE-II and OGLE-III fields toward the Galactic bulge, increasing to
357 objects the OGLE-III catalog of type~II Cepheids in the Galactic
center. Four of the newly detected RV~Tau stars belong to the RVb class,
\ie they show large-amplitude, long-period modulation of the mean
luminosity. In the updated catalog, the relative number of RV~Tau stars in
the whole sample of the Galactic bulge type~II Cepheids is similar to the Magellanic
Cloud samples.}{Stars: variables: Cepheids -- Stars: oscillations
(including pulsations) -- Stars: Population II -- Galaxy: center}

In the paper by Soszyñski \etal (2011) we presented a catalog of 335
type~II Cepheids -- BL~Her, W~Vir, and RV~Tau stars -- detected in the
OGLE-II and OGLE-III fields toward the Galactic bulge. However, during the
search for long-period variables in the same fields (Soszyñski \etal 2013)
we identified additional 22 type~II Cepheids. All of them have pulsating
periods longer than 20~days, \ie in the RV Tau star period range.

\begin{figure}[b]
\centerline{\includegraphics[width=11.4cm, bb=65 220 555 755]{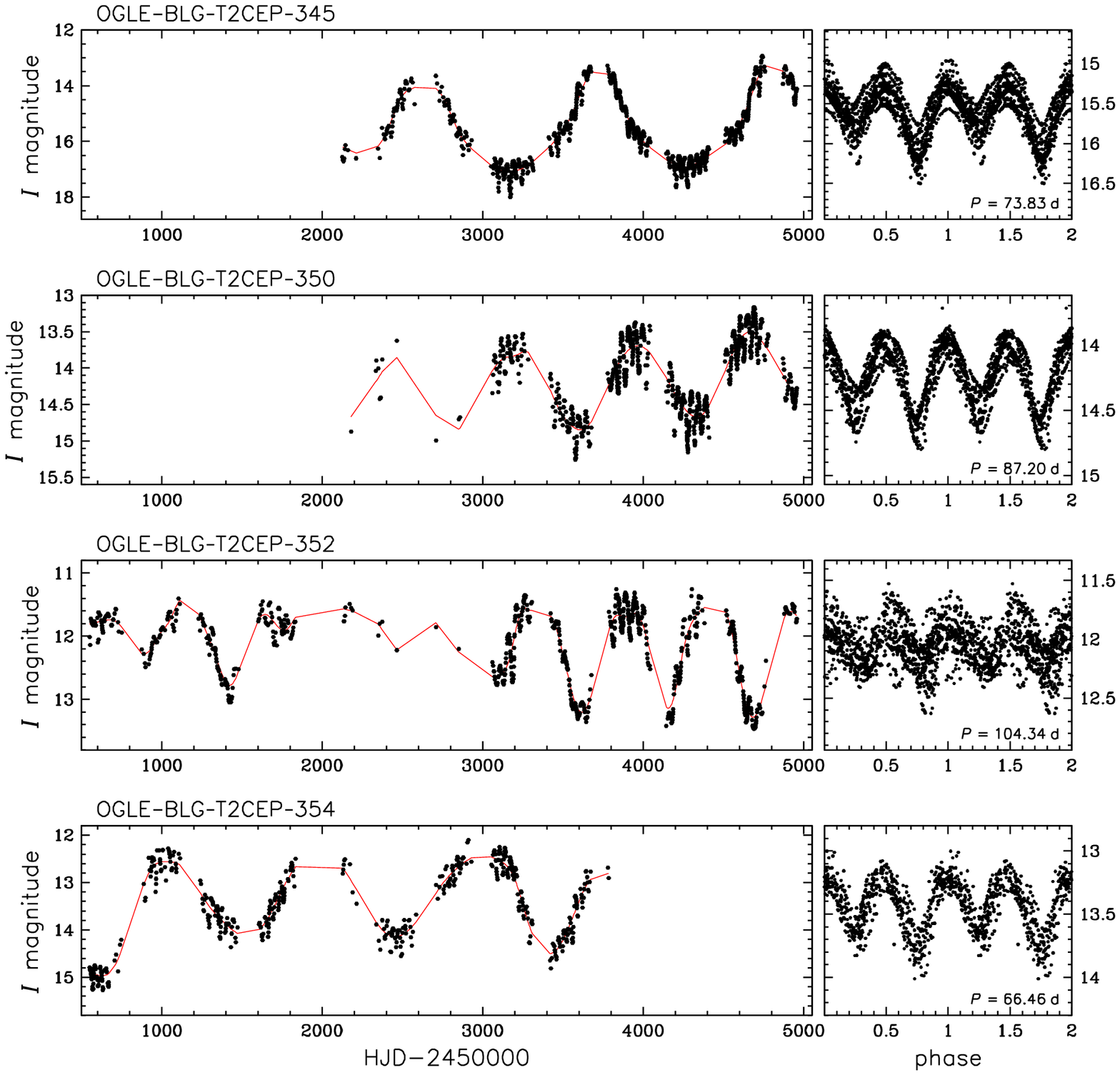}}
\FigCap{{\it I}-band light curves of RVb stars from the updated sample. {\it
Left panel} shows the unfolded light curves. Red lines are the spline functions
fitted to the light curves. {\it Right panel} presents the light curves folded
with the formal pulsation periods after subtracting the spline functions}
\end{figure}
The newly detected Cepheids were missed in the original search for a few
reasons.
\begin{enumerate}
\parskip=0pt \itemsep=1mm \setlength{\itemsep}{0.4mm}\setlength{\parindent}{-1em} \setlength{\itemindent}{0em}
\item The light curves were affected by a small number of observing points
which reduced the periodic signal-to-noise ratio below the adopted
thresholds.
\item The light curves resembled those of semiregular variables, so these
stars were initially classified as pulsating red giants. Careful
re-analysis of their intrinsic colors (dereddened with the extinction maps
by Nataf \etal 2013) and the light curve morphology showed that they are
RV~Tau stars or related to them yellow semiregular variables (SRd stars).
\item The light curves were dominated by long-term, usually periodic
changes of the mean brightness, and only the search for long-period
variables revealed these objects.
\end{enumerate}

The latter behavior in four stars is shown in Fig.~1. These are the so
called RVb stars. One object of this type -- OGLE-BLG-T2CEP-177 -- was
noticed in Soszyñski \etal (2011) with an extremely long period of the
light curve modulation of about 2800~days. The newly identified RVb stars
exhibit long periods from about 540 to 1090~days.

\begin{figure}[b]
\centerline{\includegraphics[width=9.5cm, bb=65 250 555 755]{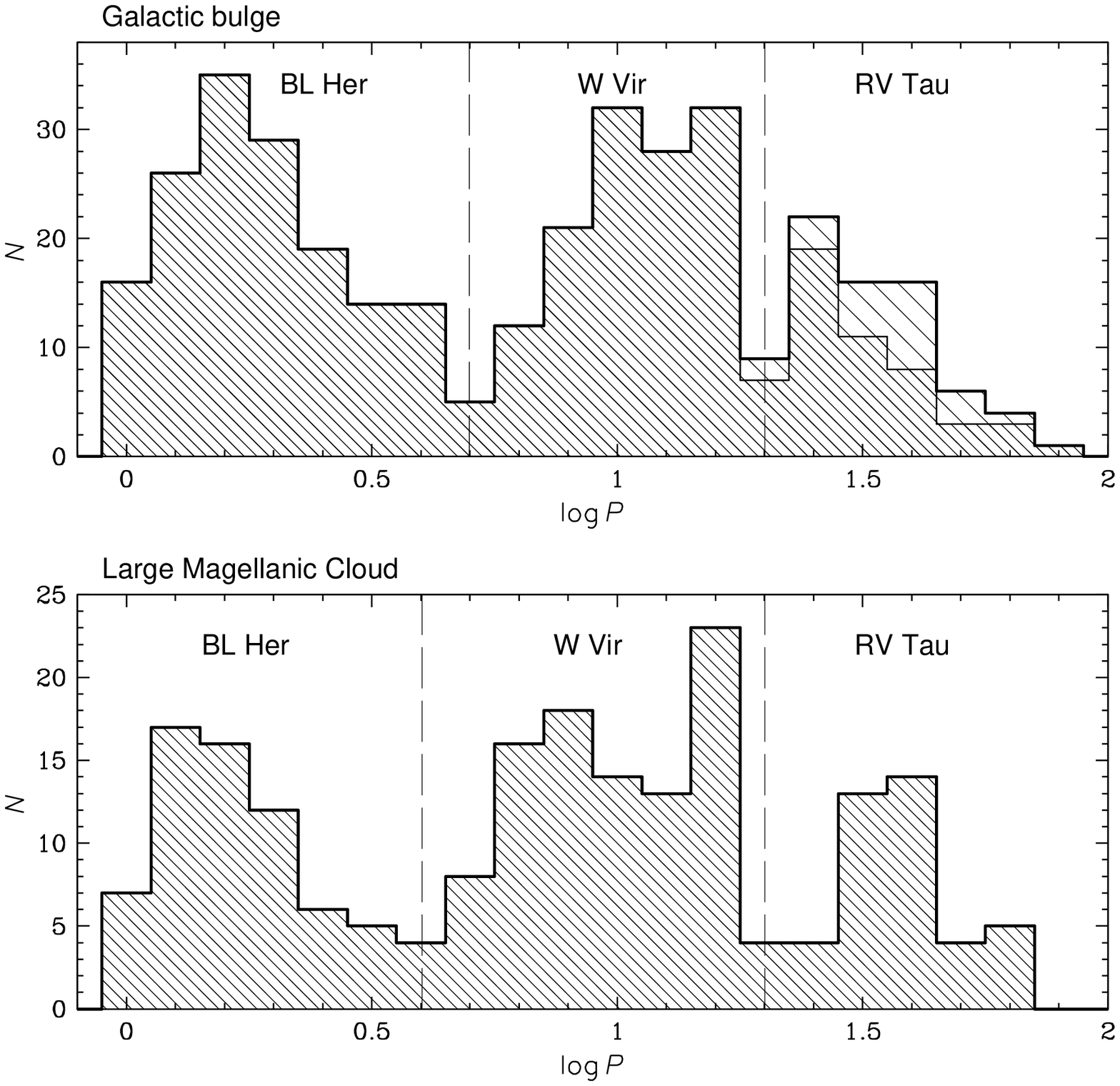}}
\FigCap{Histograms showing the period distribution of type~II Cepheids in
the Galactic bulge ({\it upper panel}) and Large Magellanic Cloud ({\it lower
panel}). The newly found RV~Tau stars in the Galactic bulge have different
shading. Vertical dashed lines show the limiting periods, used to
divide BL~Her, W~Vir and RV~Tau stars.}
\end{figure}

The OGLE-III catalog of type~II Cepheids toward the Galactic disk have been
supplemented by the new sample. The whole data set is available through the
FTP site or {\it via} the WWW interface:
\begin{center}
{\it ftp://ftp.astrouw.edu.pl/ogle/ogle3/OIII-CVS/blg/t2cep/}\\
{\it http://ogle.astrouw.edu.pl}\\
\end{center}
Note that even though our search for long-period variables covered periods
down to 5 days, we did not detect any new type~II Cepheids with periods
below 20~days (W~Vir stars). Therefore, we assess that the completeness of
our catalog of W~Vir stars toward the Galactic bulge is very high.

Three RV~Tau stars from the new sample are more than 2~mag fainter than
typical RV~Tau stars in the bulge. These objects are likely located behind
the bulge, possibly in the Sagittarius Dwarf Galaxy.

In the supplemented version of the catalog of type~II Cepheids in the
Galactic bulge the distribution of pulsation periods (Fig.~2) much more
resembles the sample from the Large Magellanic Cloud (Soszyñski \etal
2008). Now RV~Tau stars constitute 20\% of the total sample of bulge
type~II Cepheids, while in the Large and Small Magellanic Clouds this
quantity is about 21\%. This suggests similar scenarios of the late stellar
evolution in these environments.

\Acknow{This work has been supported by the Polish Ministry of Science and
Higher Education through the program ``Ideas Plus'' award No.~IdP2012 000162.
The research leading to these results has received funding from the
European Research Council under the European Community's Seventh Framework
Program\-me (FP7/2007-2013)/ERC grant agreement no.~246678.}


\begin{references}
\refitem{Nataf, D.M., \etal}{2013}{\ApJ}{~}{in press, arXiv:1208.1263}
\refitem{Soszyñski, I., Udalski, A., Szymañski, M.K., Kubiak, M., Pietrzyñski, G., Wyrzykowski, £., Szewczyk, O., Ulaczyk, K., and Poleski, R.}{2008}{\Acta}{58}{293}
\refitem{Soszyñski, I., Udalski, A., Pietrukowicz, P., Szymañski, M.K., Kubiak, M., Pietrzyñski, G., Wyrzykowski,~£., Ulaczyk,~K., Poleski,~R., and Koz³owski,~S.}{2011}{\Acta}{61}{285}
\refitem{Soszyñski, I., Udalski, A., Szymañski, M.K., Kubiak, M., Pietrzyñski, G., Wyrzykowski,~£., Ulaczyk,~K., Poleski, R., Koz³owski, S., Pietrukowicz, P., and Skowron,~J.}{2013}{\Acta}{63}{21}
\end{references}
\end{document}